\documentclass[11pt]{article}

\usepackage{amsmath}
\usepackage{geometry} 
\usepackage{graphicx} 
\usepackage{color} 
\usepackage{amssymb, amsthm} 
\usepackage[sort&compress,numbers]{natbib}
\usepackage{hyperref} 

\usepackage{booktabs}
\usepackage[table,xcdraw,dvipsnames]{xcolor}
\usepackage{soul}
\usepackage{caption}
\usepackage{subcaption}
\usepackage{lscape}

\usepackage{setspace}

\newcommand{\Var}{\text{Var}}

\newtheorem{heuristic}{Heuristic}

\newcommand{\blind}{0}

\title{A sample size heuristic for network scale-up studies}

\if1\blind
{
  \author{[Blinded for review]}
}\fi

\if0\blind
{
\author{Nathaniel Josephs$^1$, Dennis M. Feehan$^2$, and Forrest W. Crawford$^{1,3,4,5}$ \\[1em]
\normalsize 1. Department of Biostatistics, Yale School of Public Health \\
\normalsize 2. Department of Demography, University of California Berkeley \\
\normalsize 3. Department of Statistics \& Data Science, Yale University \\
\normalsize 4. Department of Ecology \& Evolutionary Biology, Yale University \\
\normalsize 5. Yale School of Management }
}\fi

\date{}

\begin{document}
\maketitle

\begin{abstract}
\noindent The network scale-up method (NSUM) is a survey-based method for estimating the number of individuals in a hidden or hard-to-reach subgroup of a general population.
In NSUM surveys, sampled individuals report how many others they know in the subpopulation of interest (e.g. ``How many sex workers do you know?'') and how many others they know in subpopulations of the general population (e.g. ``How many bus drivers do you know?'').
NSUM is widely used to estimate the size of important epidemiological risk groups, including men who have sex with men, sex workers, HIV+ individuals, and drug users.  Unlike several other methods for population size estimation, NSUM requires only a single random sample and the estimator has a conveniently simple form.
Despite its popularity, there are no published guidelines for the minimum sample size calculation to achieve a desired statistical precision.
Here, we provide a sample size formula that can be employed in any NSUM survey.
We show analytically and by simulation that the sample size controls error at the nominal rate and is robust to some forms of network model mis-specification.
We apply this methodology to study the minimum sample size and relative error properties of several published NSUM surveys.
\\[1em]
\end{abstract}

\textbf{Abbreviations}:
AIDS: acquired immunodeficiency syndrome;
ER: Erd\H{o}s-R\'{e}nyi;
ERGM: exponential random graph model; 
FSW: female sex workers; 
G-NSUM: generalized NSUM;
HIV: human immunodeficiency virus;
MMT: methadone maintenance therapy;
MSM: men who have sex with men;
NSUM: network scale-up method;
PA: preferential attachment;
SBM: stochastic block model


\section{Introduction}

Estimating the size of a hidden or hard-to-reach population is an important problem in demography, epidemiology, and public health research \citep{cheng2020estimating}.
Traditional methods, such as capture-recapture \citep{van2015estimating, karami2017estimating, paz2011how, khan2017one, robles1988application, hickman2006estimating, bouchard2007capture, bohning2004estimating, bailey1951estimating} and benchmark multiplier methods \citep{safarnejad2017population, hickman2006estimating, zhang2007advantages, zhang2007estimating} require multiple independent samples, but it can be costly to obtain multiple samples and difficult to guarantee that samples collected under different designs are independent \citep{abdul2014estimating}.  \citet{bernard1991estimating} introduced the network scale-up method (NSUM) to estimate the number of people who died in the 1985 Mexico City earthquake.  NSUM has subsequently been employed to estimate the number of men who have sex with men \citep{ezoe2012population,
wang2015application, sulaberidze2016population, guo2013estimating}, sex workers \citep{maghsoudi2014network, jami2021population, jing2018combining, guo2013estimating}, trafficked persons \citep{shelton2015proposed}, infected or at-risk individuals \citep{killworth1998estimation, killworth1998social, unaids2010guidelines, salganik2011assessing, shokoohi2012size, center2012estimating, guo2013estimating, jafarikhounigh2014size, jing2014estimating, teo2019estimating, heydari2019evaluation, haghdoost2015application, sajjadi2018indirect}, drug users \citep{kadushin2006scale, nikfarjam2016national, sheikhzadeh2016comparing, guo2013estimating, maghsoudi2014network}, prisoners \citep{zheng2006many}, victims of disasters \citep{bernard2001estimating}, abortions \citep{rastegari2014estimating}, choking incidents in children \citep{snidero2007use}, religious individuals \citep{yang2017estimating}, and people in one's personal network \citep{mccormick2010many, narouee2020estimating}.

As \citet{maltiel2015estimating} summarized, ``NSUM is based on the idea that for all individuals, the probability of knowing someone in a given subpopulation is the size of that subpopulation divided by the overall population size.''
In the basic NSUM \citep{bernard2010counting, mccormick2020network, laga2021thirty, habecker2017you}, investigators obtain a single random sample of individuals in the general population, who are not necessarily 
members of the hidden population.
Each respondent reports the number of others they know in the general population (or the number they know in several subgroups of the general population so that the number of others they know in the general population may be estimated) and also the number they know in the target population.
The ratio of these average counts is multiplied by the known size of the general population to estimate the size of the target population.

Population size estimation using NSUM relies on several assumptions related to homogeneity of the underlying network of acquaintanceships and the accuracy of reported counts.
Several variants and generalizations of NSUM extend the method to accommodate more flexible assumptions using complex estimators.
For example, \citet{mccormick2007adjusting} provide an adjustment for recall bias, in which the number of contacts is underestimated in large groups and overestimated in small groups. 
\citet{habecker2015improving} and \citet{feehan2016generalizing} formalize the incorporation of unequal sampling weights for surveyed individuals into NSUM.
\citet{maltiel2015estimating} propose a Bayesian approach that models recall bias, transmission error, and barrier effects directly.
\citet{feehan2016generalizing} introduce a method that generalizes NSUM (called G-NSUM) by recognizing that in an undirected network, total in-degrees and out-degrees must be equal, a fact that can be exploited when two samples are obtainable.
The G-NSUM was further generalized for venue-based sampling \citep{verdery2019estimating}.
While these methods offer improved and more efficient estimates, the classic NSUM method remains widely used due to its simplicity. 

As NSUM surveys grow in popularity, practitioners need guidance on how to design NSUM studies to achieve accurate and reliable results.  Researchers have noted that population-size estimates from NSUM can have high variance \citep{bernard2010counting, maghsoudi2014network, mccormick2010many, salganik2011assessing}.
This may be due in part to the lack of a coherent framework that guides investigators in choosing a sample size for empirical studies.  To address this gap, we present a sample size heuristic that enables researchers to calculate the number of respondents needed to estimate the size of a hidden population at a given relative error.
We investigate the properties of the sample size heuristic and analyze its performance under network model mis-specificiation. 
Finally, we perform a retrospective sample size analysis for several existing NSUM studies and conclude with recommendations for empirical NSUM study design. 



\section{Setting and NSUM estimator} \label{sec:setting}

Consider a set $V$ of individuals of known size $M = \vert V\vert$ called the \textit{general population}. 
A subset of individuals $U \subset V$ of size $N = \vert U\vert$ comprise the \textit{hidden population}; we wish to estimate $N$. 
NSUM relies on information about the network of relationships between individuals. The meaning of these relationships varies according to the type of study, but may include social, epidemiological, or communication relationships.
Let $G = (V,~E)$ represent the general population network, where $V$ is the set of individuals in the general population and $E$ is the set of relevant relationships between population members. We assume $G$ is undirected and simple, that is, contains no parallel edges or self-loops.
Then for all $i, j \in V$, we have $(i, j) \in E$ whenever $i$ and $j$ share the relationship of interest.
Next, define $d_i^u = \vert\{(j,i)\in E,~ j\in U\}\vert$ to be the \textit{number of connections to the hidden population} of person $i$, that is, the number of links between person $i$ and members of the hidden population $U$.
Finally, define the \textit{personal network size} of person $i$ -- which is also called $i$'s \textit{degree} -- as $d_i = \vert\{(j,i)\in E,~ j\in V\}\vert$, the number of edges incident to person $i$ in $G$.
For simplicity, we assume below that $d_i$ is directly observed, but in practice NSUM studies often estimate $d_i$ using the known
population method \cite{killworth1998social, killworth1998estimation} or the summation method~\cite{mccarty2001comparing}.

The statistical performance of the NSUM estimator relies on assumptions about the distribution of personal network sizes $d_i$ and reported connections to the hidden population $d_i^u$ that imply certain global features of the underlying general population network.
In many cases, these assumptions correspond to a particular random graph model.
\citet{laga2021thirty} provides a summary of network distributional assumptions in the NSUM literature.
In what follows, we assume the general population network has Erd\H{o}s-R\'{e}nyi distribution \citep{erdos1959random}. 
This framework can be extended to any vertex-exchangeable graph model \citep{diaconis2007graph}, and we illustrate in Section \ref{sec:simulations}, with details in Appendix \ref{appendix:simulations}, that the sample size heuristic is robust to violations of vertex-exchangeability.

Under an Erd\H{o}s-R\'{e}nyi model for the network $G$, edge relationships exist between individuals in $V$, and between individuals in $V$ and $U$, with probability $p$ independently of other relationships.
Consequently, the following \textit{marginal degree models} for person $i$ hold:
\begin{equation}\label{eq:marg_bin}
    \begin{split}
        d_i &\sim \text{Binomial}(M-1,~p) \\
        d_i^u &\sim \text{Binomial}(N,~p) \enskip .
    \end{split}
\end{equation}
It follows that
\begin{equation*}
    \begin{split}
        \mathbb{E}[d_i] &= (M-1)p \approx Mp \\
        \mathbb{E}[d_i^u] &= Np \enskip .
    \end{split}
\end{equation*}
If $i$ is a member of the target population, then $d_i^u \sim \text{Binomial}(N-1,~p)$ and $\mathbb{E}[d_i^u] = (N-1)p \approx Np$.

Solving for $p$ gives
\begin{equation}\label{eq:nsum_model}
    p \approx \mathbb{E}[d_i^u]~/~N = \mathbb{E}[d_i]~/~M \enskip .
\end{equation}
This yields an immediate formula for $N$ in terms of the expected general and hidden population degrees:
\begin{equation}\label{eq:N}
    N \approx M \cdot \mathbb{E}[d_i^u]~/~\mathbb{E}[d_i] \enskip .
\end{equation}

To estimate the hidden population size $N$, consider a random sample of $n$ individuals from $V$, each of whom reports how many individuals they know in 
different subgroups of the general population (which allows us to estimate $d_i$) 
and how many individuals they know in the hidden population (which is $d_i^u$).
Using this information, NSUM estimates both of the expectations in \eqref{eq:N} empirically by averages.
Then a method of moments equation for \eqref{eq:nsum_model} is
\begin{equation*}
    N \cdot \frac{1}{n}\sum_{i=1}^n d_i = M \cdot \frac{1}{n}\sum_{i=1}^n d_i^u.
\end{equation*}
Solving for $N$, we arrive at the classic NSUM estimator \citep{killworth1998estimation},
\begin{equation}\label{eq:nsum}
    \widehat{N} = M \cdot \frac{\sum_{i=1}^n d_i^u}{\sum_{i=1}^n d_i} \enskip .
\end{equation}



\section{Sample size heuristic}
\label{sec:methods}

Investigators conducting an NSUM survey must determine the minimum number of samples to collect in order to obtain a suitably precise estimate of the hidden population size $N$.
To do this, investigators must first specify how precise they want their estimates to be.
This means choosing a value for two parameters:
(i) the \textit{confidence level}, $1 - \alpha$; and
(ii) the \textit{relative margin of error}, $\varepsilon$.
In many applied studies, researchers choose to use 95\% confidence intervals to express sampling uncertainty
in their estimates, meaning that $\alpha = 0.05$.
The relative margin of error, $\varepsilon$, describes how close, in relative terms, the estimate is
expected to be to the population value \cite{valliant2013practical}. 
For example, a relative margin of error of 10\% ($\varepsilon = 0.1$) and a confidence level of 95\% ($\alpha = 0.05$) means that researchers want a sample size that produces estimates within 10\% of the true value
at least 95\% of the time (that is, in at least 95\% of repeated samples).

Formally, the minimum sample size is the smallest $n$ such that
\begin{equation}\label{eq:error}
    \Pr\left(\frac{\vert\widehat{N}-N\vert}{N} < \varepsilon \right) \ge 1-\alpha \enskip ,
\end{equation}
where $\vert\widehat{N}-N\vert/N$ is the relative absolute error of the estimate. 
In other words, investigators seek the smallest sample size that ensures the relative error is small with high probability. 


\begin{heuristic}\label{prop:samp_size}
    Let the relative margin of error be $\varepsilon > 0$, the confidence level be $1 - \alpha$ for $0 < \alpha < 1$, the prevalence of the hidden population be $q = N/M$, and the average personal network size be $\overline{d} = M^{-1} \sum_{i=1}^M d_i$.
    The minimum sample size $n$ needed to satisfy \eqref{eq:error} is then
    \begin{equation}\label{eq:samp_size}
        n = \left\lceil \frac{z_{\alpha/2}^2}{\varepsilon^2}\cdot\frac{1}{q}\cdot\Big(\frac{1}{\overline{d}} - \frac{1}{M}\Big) \right\rceil \enskip ,
    \end{equation}
    where $\lceil \cdot \rceil$ is the ceiling operator and $z_{\alpha/2}$ is the $1-\alpha/2$ quantile of the standard normal distribution.
\end{heuristic}

In order to use the sample size formula \eqref{eq:samp_size}, researchers must specify the values for three population
quantities: 
(i) the size of the general population, $M$;
(ii) the prevalence of the hidden population, $q = N/M$;
and (iii) the average size of respondents' personal networks, $\overline{d}$.
Several properties of the minimum sample size can be deduced immediately from \eqref{eq:samp_size}, and
these properties can be helpful in guiding the choice of these three parameters.
First, when the general population size $M$ is large, the contribution of $M$ to the minimum sample size in \eqref{eq:samp_size} is negligible. 
Second, the consequences of error in estimates of $q$ and $\overline{d}$ are evident.
If the prevalence of the hidden population $q$ is under-estimated, the minimum sample size increases; if $q$ is over-estimated, the minimum sample size increases.
Therefore, it is preferable to over-estimate the prevalence to ensure a conservative sample size.
If the average population degree $\overline{d}$ is under-estimated, the minimum sample size increases; if $\overline{d}$ is over-estimated, the minimum sample size decreases. Therefore it may be preferable for investigators seeking a conservative sample size to use an under-estimate of the average general population degree.

\begin{figure} 
    \centering
    \includegraphics[width=\textwidth]{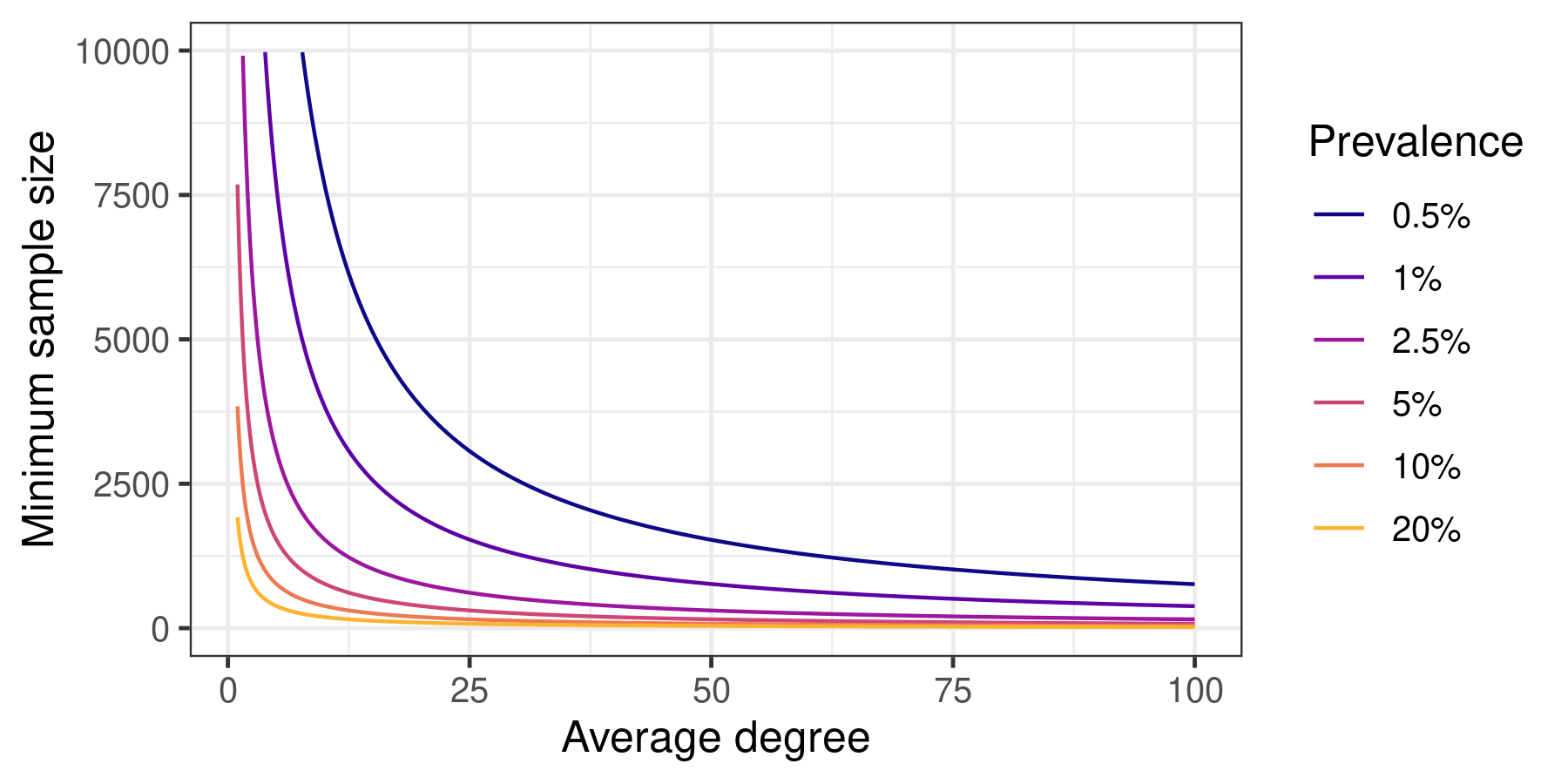}
    \caption{Sample size as a function of prevalence $(q)$ and average degree $(\overline{d})$ for $M = 10,000$. We truncate the minimum sample size at the population size $M$.}
    \label{fig:dependence}
\end{figure}

Figure \ref{fig:dependence}, shows the relationship between prevalence and average degree in terms of the minimum sample size for 
a confidence level of 95\% ($\alpha = 0.05$), a relative margin of error of 10\% ($\varepsilon = 0.1$), and a general population size of $M = 10,000$.
For example, when $\alpha = 0.05$, we have $z_{\alpha/2}\approx 2$, and thus if the relative margin of error is $\varepsilon=0.1$, the prevalence is $q = 0.1$ and the average degree is $\overline{d} = 10$, then the minimum sample size from \eqref{eq:samp_size} is approximately
\begin{equation*}
    n \approx \frac{2^2}{0.1^2} \cdot \frac{1}{0.1} \cdot \frac{1}{10} = 4 * 100 = 400 \enskip .
\end{equation*}

Heuristic \ref{prop:samp_size} is appropriate when subjects are sampled uniformly at random from the  population.  In practice, NSUM surveys may employ more complex sampling designs using multi-stage sampling, clustering, and stratification.
These complex design features affect the variance of estimates.
One common approach to modifying the sample size heuristic in \eqref{eq:samp_size} is through the use of a \textit{design effect} for quantifying the impact of a complex sampling design on an estimate.
It is defined as the ratio of the variances of an estimator with a given sampling design to a simple random sample; if $\widehat{N}_w$ is an NSUM estimator with a complex sampling design, then
\begin{equation}\label{eq:d_eff}
    D_\textit{eff} = \frac{\widehat{N}_w}{\widehat{N}} \enskip .
\end{equation}


%

If the design effect is known \emph{a priori}, then the minimum sample size calculation from Heuristic \ref{prop:samp_size} can be multiplied by it to obtain an adjusted estimate.

\begin{heuristic}\label{corollary}
    Let $D_\textit{eff}$ be the known design effect for a given sampling scheme.
    The minimum sample size $n$ needed to achieve \eqref{eq:error} is approximately
    \begin{equation}
        n =\left\lceil \frac{z_{\alpha/2}^2}{\varepsilon^2}\cdot\frac{1}{q}\cdot \frac{1}{\overline{d}} \cdot D_\textit{eff} \right\rceil \enskip .
    \end{equation}
\end{heuristic}

Alternatively, once a sample has been collected, an \textit{effective sample size} can be computed as $n_\textit{eff} =~ \frac{n}{D_\textit{eff}}$.



\section{Results}\label{sec:results}


\subsection{Simulation studies}\label{sec:simulations}

We first assess the relative error and coverage rates using the minimum sample size by simulation. 
We employ a factorial design that varies the population size, the prevalence, the nominal levels, and the underlying population graph model.
We include i) an Erd\H{o}s-R\'{e}nyi network \citep{erdos1959random}, ii) an exponential random graph \citep{robins2007introduction}, iii) a preferential attachment network \citep{barabasi1999emergence}, iv) a stochastic block model \citep{holland1983stochastic}, and v) a small-world network \citep{watts1998collective}.
Each graph model has the same expected density around $10\%$, which allows us to isolate the effect of the graph topology and assess the robustness to network model mis-specification, i.e. how well our minimum sample size calculation works when the underlying population graph implies degree models that violate \eqref{eq:marg_bin}.

The details and results are left to Appendix \ref{sec:simulations}, but the conclusions are as follows.
The average relative error is below the tolerated level $\varepsilon$ for all values of $M$ and $q$ when the underlying graph model is ER, ERGM, and small-world.
For low prevalence, PA and SBM have average relative errors that exceed the tolerance level, but this is mitigated as prevalence and population size increase.
Similarly, the average coverage is conservative for ER, ERGM, and small-world across different nominal levels $\alpha$, whereas it suffers for small populations with low prevalence for PA and SBM.
As expected, the error bars are larger for larger values of $\alpha$.
Finally, the relative error and coverage drift smoothly away from the desired levels as the underlying graph model deviates from the assumption of vertex-exchangeability. 

\subsection{Case studies}

We now conduct a retrospective sample size analysis of published empirical NSUM surveys.  Each published study reports the population size $M$, the sample size $n$ employed in the study, and the population size estimate $\widehat{N}$.  To derive information about the marginal degree model, we assume that the degree reports $d_i$ and $d_i^u$ follow \eqref{eq:marg_bin}, 
\begin{equation} \label{eq:sim}
    \begin{split}
        d_i &\sim \text{Binomial}(M,~\hat{d}_i/M) \\
        d_i^u &\sim \text{Binomial}(\widehat{N},~\hat{d}_i^u/\widehat{N}) \enskip ,
    \end{split}
\end{equation}
where $\hat{d}_i$, $\hat{d}_i^u$, and $\widehat{N}$ are the estimates from the study.
In other words, we take the published estimate $\widehat{N}$ as the true hidden population size $N$.
From this information, we calculate the minimum sample sizes $n$ according to \eqref{eq:samp_size} using the implied average population degree and prevalence given the published value of the general population size $M$.
We use a relative error tolerance of 10\% ($\varepsilon = 0.1$) and 95\% confidence level ($\alpha = 0.05$).

Using these parameters, we compute 10,000 Monte Carlo NSUM estimates of the target population size.
That is, in each replicate, we sample $n$ degree reports $d_1, \ldots, d_n$ and $d_1^u, \ldots, d_n^u$ following the Binomial models in \eqref{eq:sim}, and use them to calculate the NSUM estimate of $N$. 
For each replicate, we compute the relative error of our NSUM estimate compared to the published estimate in the study, and report the Monte Carlo average relative error over all 10,000 replicates, which we denote RelErr.
We apply this retrospective sample size calculation to seven published NSUM studies.

First, we revisit a study by \citet{killworth1998estimation} to estimate the number of HIV+ individuals in 1994 in the US, around the height of the AIDS crisis.
Using a telephone survey, the authors randomly sampled $n=1,554$ members of the US population.
Despite the above-average response rate, the authors recognized the potential for some non-response bias.
However, they do not discuss why this sample size was chosen, but they report that the survey cost \$6.50 per respondent and took 10 minutes to conduct, which suggests there may have been resource constraints.
One of the major contributions of this study is an improved technique for estimating personal network sizes, along with one of the first reports of the distribution of network sizes for a random sample of the US population.
Using these estimates, the authors concluded that a 95\% confidence interval for the number of HIV+ individuals was $\hat{N} = 800,000 \pm 43,000$, which was in line with estimates from the CDC on seroprevalence.

Second, we look at two applications of NSUM to estimate the number of sex workers in China \citep{guo2013estimating, jing2018combining}.
Chongqing, the largest province in China, had its first reported HIV case in 1993.
By 2011, there were nearly 12,000 reported cases, yet population estimates of key affected populations were then unknown.
Citing the need for targeted interventions and resource allocation, \citet{guo2013estimating} employed a multistage random sample of 2,957 individuals in order to obtain NSUM estimates of key populations including female sex workers (FSW).
Using their survey results, the authors estimated $\overline{d} = 311$, which had not previously been estimated in Chongqing.
This led to a 95\% confidence interval of $\hat{N} = 31,576 \pm 1,980$ for the number of FSW.
They also found that sex workers have a high respect factor, but, while they argue that this suggests their NSUM estimates are less likely to be underestimated, their estimate when adjusting for respect factor was 10\% lower.
Although this is an important study, being one of the first applications of NSUM in China, the authors note several limitations that ultimately bring additional uncertainty to the conclusions.

More recently, \citet{jing2018combining} incorporated a randomized response technique to an NSUM survey in 2012 of 7,964 individuals in Taiyuan, China.
The authors estimate the FSW population to be $\hat{N} =~3,866$, which was similar to the more expensive multiplier methods and consistent with previously reported estimates of sex worker prevalence in Asia. The authors argue that this demonstrates the the appropriateness of their adjusted NSUM approach.

Next, we look at two applications of NSUM to estimate the number of men who have sex with men (MSM) \citep{ezoe2012population, wang2015application}.
Since 2008, the number of HIV cases in Japan has risen constantly, with 89\% of new cases attributed to MSM.
However, prior to 2012, the size of the MSM population in Japan had not been estimated ``in a rigorous manner" \citep{ezoe2012population}.  
To address this, \citet{ezoe2012population} employed the first internet-based NSUM estimator.
The authors surveyed 1,500 individuals who were registered to \textit{Intage}, an internet-research agency.
They estimated the MSM prevalence among the male population to be 2.87\%, which was comparable to previous studies using the direct-estimation method.
These results suggest that internet surveys can be combined with NSUM for an even quicker and lower-cost method, especially in stigmatized populations.

\citet{wang2015application} are also interested in estimating the number of MSM, but in Shanghai, China.
Instead of an internet-based survey, the authors conducted a community-based survey across 19 districts in Shanghai consisting of 3,907 participants.
Using NSUM, the authors report a 95\% confidence interval for $\hat{N}$ as $36,354 \pm 7,865$, but they note that ``we did not take the sample design into consideration when performing variance estimation."

Finally, we consider two applications of NSUM to quantify drug use \citep{habecker2015improving, heydari2019evaluation}.
In the first, \citet{heydari2019evaluation} are interested in estimating the number of individuals using methadone maintenance therapy (MMT) in Kerman, Iran.
Such data, which was previously not available, is needed to assess the effectiveness of MMT.
The authors used two cross-sectional studies with multi-stage sampling to recruit 2,550 individuals.
Using NSUM, they report $\hat{N} = 5,289$.
Furthermore, they were able to use this to estimate the treatment failure ratio, which was a novel application of NSUM and further evidence of its widespread applicability.

Similarly, \citet{habecker2015improving} use NSUM to estimate the number of heroin users in Nebraska in the past 30 days.
In 2014, the authors received 550 completed mail surveys, which included additional questions that allowed the authors to compute and incorporate sampling weights into an improved NSUM estimator.
Such improvements are particularly important when knowledge of the scope of the problem is time-sensitive.
Ultimately, this yielded an estimate of a 95\% confidence interval of $\hat{N} = 368 \pm 89$ for the number of heroin users in the past 30 days, which may be a proxy for the number of current heroin users.

\begin{table} 
\centering
\resizebox{\textwidth}{!}{%
\begin{tabular}{@{}|l|r|r|r|r|r|r|r|@{}}
\toprule
\multicolumn{1}{|c|}{\textbf{Hidden population}} & \multicolumn{1}{c|}{$n$} & \multicolumn{1}{c|}{$M$} & \multicolumn{1}{c|}{$\widehat{N}$} & \multicolumn{1}{c|}{$\overline{d}_i$} & \multicolumn{1}{c|}{$\overline{d}_i^u$}
& \multicolumn{1}{c|}{RelErr} & \multicolumn{1}{c|}{$n_\textit{min}$} \\

\midrule

\rowcolor[HTML]{EFEFEF} Heroin users in Nebraska \citep{habecker2015improving} & 550 & 1,879,321 & 368 & 604 & 0.118 & 0.02 & 3,383 \\

FSW in Taiyuan, China \citep{jing2018combining} & 7,964 & 3,454,927 & 3,866 & 137 & 0.15 & 0.03 & 2,610 \\

\rowcolor[HTML]{EFEFEF} MMT users in Kerman, Iran \citep{heydari2019evaluation} & 2,550 & 611,401 & 5,289 & 235 & 2.03 & 0.01 & 197 \\

FSW in Chongqing, China \citep{guo2013estimating} & 2,957 & 28,000,000 & 31,576 & 311 & 0.077 & 0.78 & 1,141 \\

\rowcolor[HTML]{EFEFEF} MSM in Shanghai, China \citep{wang2015application} & 3,907 & 24,000,000 & 36,354 & 236 & 0.159 & 0.56 & 1,119 \\

HIV+ individuals in US \citep{killworth1998estimation} & 1,554 & 250,000,000 & 800,000 & 286 & 0.91 & 0.02 & 438 \\

\rowcolor[HTML]{EFEFEF} MSM in Japan \citep{ezoe2012population} & 1,500 & 62,348,977 & 1,789,416 & 174 & 5.09 & 0.02 & 81 \\





\bottomrule
\end{tabular}%
}
\caption{Retrospective analysis for average relative error over 10,000 Monte Carlo estimates and minimum sample sizes from \eqref{eq:samp_size}.
The population size and average degrees are estimates from the given NSUM study.}
\label{tab:retro}
\end{table}

\begin{figure} 
    \centering
    \includegraphics[width=\textwidth]{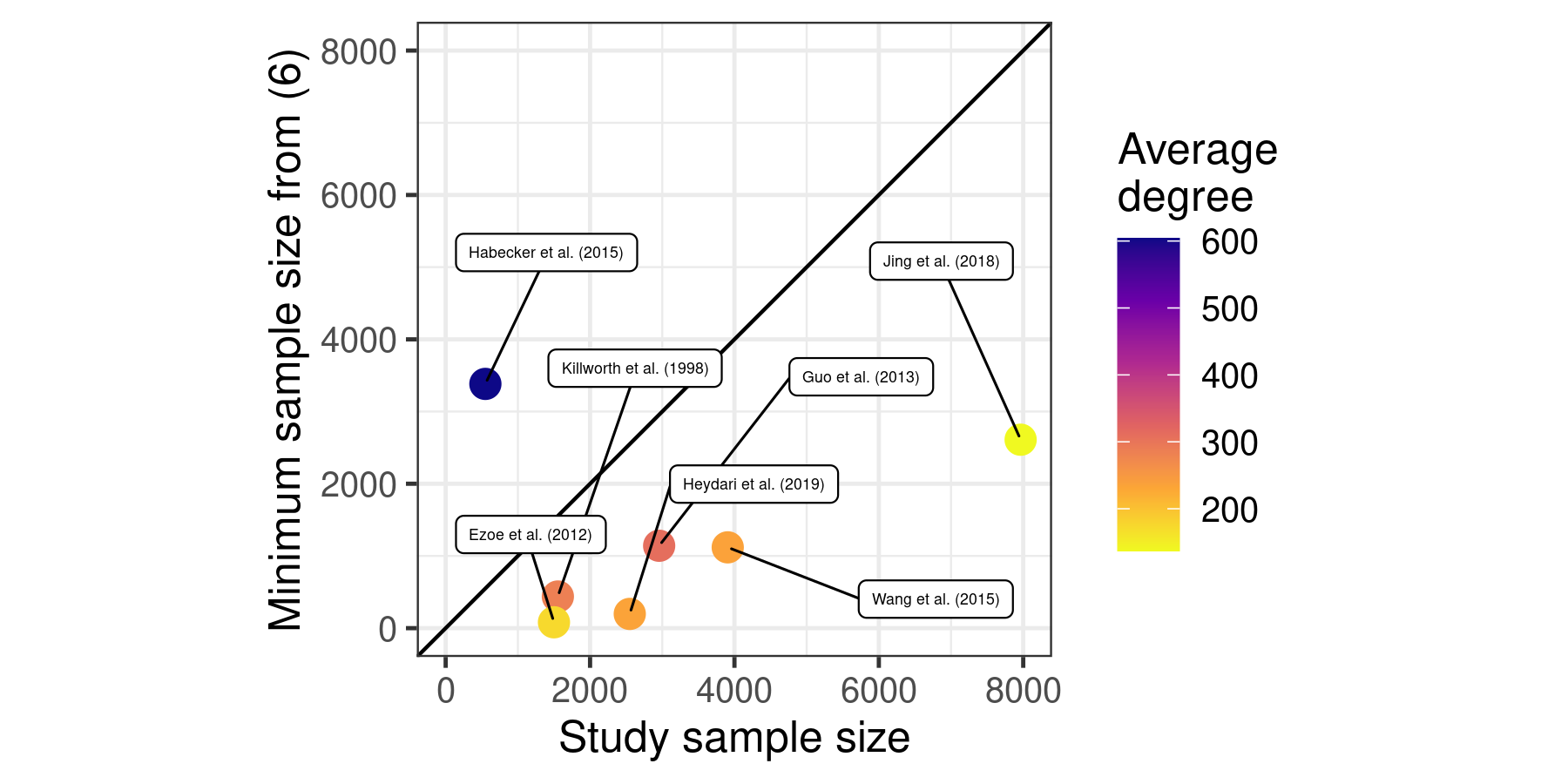}
    \caption{Study sample size compared to the minimum sample size from \eqref{eq:samp_size} calculated based on each study's reported estimates of prevalence and average network size.
    Most of the points lie below the line $y = x$, suggesting that the actual sample size was larger than necessary to achieve 10\% precision with 95\% confidence.}
    \label{fig:sample_size}
\end{figure}

Table \ref{tab:retro} shows the results of the retrospective analysis on these seven studies. Figure \ref{fig:sample_size} compares the actual study sample sizes and the retrospectively calculated minimum sample sizes using \eqref{eq:samp_size}. 
While most of the studies have an average relative error below $0.05$, \citet{guo2013estimating} and \citet{wang2015application} may have much larger relative errors.
In both of these cases, the study sample size is larger than our minimum sample size, but the average hidden population degree is much smaller than the average general population degree times the prevalence.
This suggests the topology of the hidden population network differs from that of the general population network, which was a specific limitation listed by \citet{guo2013estimating}.


\section{Discussion}\label{sec:discussion}


In this paper, we have presented a simple heuristic for the minimum sample size that controls relative error \eqref{eq:samp_size}.
The heuristic is easy to employ and theoretically justified under the assumption that the general population network follows the Erd\H{o}s-R\'{e}nyi random graph model.
Furthermore, simulations show that the sample size heuristic is robust to a variety of deviations from this idealized setting.

We also demonstrated how published NSUM studies have sample sizes that scale similarly to this minimum sample size, though most have used a sample size larger than our retrospectively calculated minimum sample size, implying that they could have saved resources by administering surveys to fewer respondents.
That said, since the minimum sample sizes are functions of the reported estimates of prevalence and average network degrees, which are all quite high, it is possible that this is a propagation of biased estimates.
On the other hand, one of the studies has $n < n_\textit{min}$, which implies they may have low precision in their estimates of the sizes of the populations of interest.
In general, practitioners appear to be intuiting our heuristic as many of the study sample sizes are on the same order of magnitude as our minimum sample sizes.

In order to use the formula \eqref{eq:samp_size}, investigators must specify values for $M$,
$q$, and $\overline{d}$.  
In some cases, estimates of $q$ and $\overline{d}$ may be based on the results of previous studies.
But, if little existing information about the hidden population is available, our results
show that it will be conservative to err on the side of assuming that the hidden population
is relatively low in prevalence (low $q$) and that personal network sizes are relatively
small (low $\overline{d}$).

While our heuristic should help practitioners design future NSUM studies, there are several limitations to our analysis.
First, the relative error controlled for in Heuristic \ref{prop:samp_size} may not provide the desired absolute precision.
Second, the normal approximation of our standardized estimator $\frac{\widehat{N} - N}{\sqrt{v_n}}$ may be invalid for small samples.
Third, sample size formula may differ for more complex sampling designs and the design effects in Heuristic \ref{corollary} for adjusting the minimum sample size may be unknown.
Finally, non-sampling error, for instance from imperfect reporting, may dominate statistical error leading to invalid estimates even for $n \approx M$ in some cases.


\textbf{Acknowledgements}

\if1\blind
{
  \author{[Blinded for review]}
}\fi

\if0\blind
{
We are grateful to Si Cheng, Jinghao Sun, and Clifford Zinnes for helpful comments.
This work was supported by the Eunice Kennedy Shriver National Institute of Child Health and Development (1DP2HD091799-01 and P2C HD 073964 via the Berkeley Population Center).
DMF also thanks the Berkeley Center for the Economics and Demography of Aging, which is funded by the National Institute on Aging (5P30AG012839).
}\fi


\bibliography{nsum_sample_size}

\begin{thebibliography}{63}
\providecommand{\natexlab}[1]{#1}
\providecommand{\url}[1]{\texttt{#1}}
\expandafter\ifx\csname urlstyle\endcsname\relax
  \providecommand{\doi}[1]{doi: #1}\else
  \providecommand{\doi}{doi: \begingroup \urlstyle{rm}\Url}\fi

\bibitem[Cheng et~al.(2020)Cheng, Eck, and Crawford]{cheng2020estimating}
Si~Cheng, Daniel~J Eck, and Forrest~W Crawford.
\newblock Estimating the size of a hidden finite set: large-sample behavior of
  estimators.
\newblock \emph{Statistics Surveys}, 14:\penalty0 1--31, 2020.

\bibitem[van~der Heijden et~al.(2015)van~der Heijden, de~Vries, B{\"o}hning,
  and Cruyff]{van2015estimating}
Peter~GM van~der Heijden, Ieke de~Vries, Dankmar B{\"o}hning, and Maarten
  Cruyff.
\newblock Estimating the size of hard-to-reach populations using
  capture-recapture methodology, with a discussion of the {International Labour
  Organization’s} global estimate of forced labour.
\newblock In \emph{Forum on Crime and Society}, volume~8, pages 109--136.
  United Nations Publications, 2015.

\bibitem[Karami et~al.(2017)Karami, Khazaei, Poorolajal, Soltanian, and
  Sajadipoor]{karami2017estimating}
Manoochehr Karami, Salman Khazaei, Jalal Poorolajal, Alireza Soltanian, and
  Mansour Sajadipoor.
\newblock Estimating the population size of female sex worker population in
  {T}ehran, {I}ran: Application of direct capture--recapture method.
\newblock \emph{AIDS and Behavior}, 27\penalty0 (8):\penalty0 1--7, 2017.

\bibitem[Paz-Bailey et~al.(2011)Paz-Bailey, Jacobson, Guardado, Hernandez,
  Nieto, Estrada, and Creswell]{paz2011how}
G~Paz-Bailey, JO~Jacobson, ME~Guardado, FM~Hernandez, AI~Nieto, M~Estrada, and
  J~Creswell.
\newblock How many men who have sex with men and female sex workers live in
  {E}l {S}alvador? {U}sing respondent-driven sampling and capture-recapture to
  estimate population sizes.
\newblock \emph{Sexually Transmitted Infections}, 87\penalty0 (4):\penalty0
  279--282, 2011.

\bibitem[Khan et~al.(2018)Khan, Lee, and Dombrowski]{khan2017one}
Bilal Khan, Hsuan-Wei Lee, and Kirk Dombrowski.
\newblock One-step estimation of networked population size with anonymity using
  respondent-driven capture-recapture and hashing.
\newblock \emph{PLoS One}, 13\penalty0 (4):\penalty0 e0195959, 2018.

\bibitem[Robles et~al.(1988)Robles, Marrett, Clarke, and
  Risch]{robles1988application}
Sylvia~C. Robles, Loraine~D. Marrett, E.~Aileen Clarke, and Harvey~A. Risch.
\newblock An application of capture-recapture methods to the estimation of
  completeness of cancer registration.
\newblock \emph{Journal of Clinical Epidemiology}, 41\penalty0 (5):\penalty0
  495--501, 1988.

\bibitem[Hickman et~al.(2006)Hickman, Hope, Platt, Higgins, Bellis, Rhodes,
  Taylor, and Tilling]{hickman2006estimating}
Matthew Hickman, Vivian Hope, Lucy Platt, Vanessa Higgins, Mark Bellis, Tim
  Rhodes, Colin Taylor, and Kate Tilling.
\newblock Estimating prevalence of injecting drug use: A comparison of
  multiplier and capture--recapture methods in cities in {E}ngland and
  {R}ussia.
\newblock \emph{Drug and Alcohol Review}, 25\penalty0 (2):\penalty0 131--140,
  2006.

\bibitem[Bouchard(2007)]{bouchard2007capture}
Martin Bouchard.
\newblock A capture-recapture model to estimate the size of criminal
  populations and the risks of detection in a marijuana cultivation industry.
\newblock \emph{Journal of Quantitative Criminology}, 23\penalty0 (3):\penalty0
  221--241, 2007.

\bibitem[B{\"o}hning et~al.(2004)B{\"o}hning, Suppawattanabodee, Kusolvisitkul,
  and Viwatwongkasem]{bohning2004estimating}
Dankmar B{\"o}hning, Busaba Suppawattanabodee, Wilai Kusolvisitkul, and Chukiat
  Viwatwongkasem.
\newblock Estimating the number of drug users in {B}angkok 2001: A
  capture-recapture approach using repeated entries in one list.
\newblock \emph{European Journal of Epidemiology}, 19\penalty0 (12):\penalty0
  1075, 2004.

\bibitem[Bailey(1951)]{bailey1951estimating}
Norman T.~J. Bailey.
\newblock On estimating the size of mobile populations from recapture data.
\newblock \emph{Biometrika}, 38\penalty0 (3/4):\penalty0 293--306, 1951.

\bibitem[Safarnejad et~al.(2017)Safarnejad, Nga, and
  Son]{safarnejad2017population}
Ali Safarnejad, Nguyen~Thien Nga, and Vo~Hai Son.
\newblock Population size estimation of men who have sex with men in {Ho Chi
  Minh City} and {Nghe An} using social app multiplier method.
\newblock \emph{Journal of Urban Health}, 94\penalty0 (3):\penalty0 339--349,
  2017.

\bibitem[Zhang et~al.(2007{\natexlab{a}})Zhang, Wang, Lv, Su, Liu, Shen, and
  Bi]{zhang2007advantages}
D.~Zhang, L.~Wang, F.~Lv, W.~Su, Y.~Liu, R.~Shen, and P.~Bi.
\newblock Advantages and challenges of using census and multiplier methods to
  estimate the number of female sex workers in a {C}hinese city.
\newblock \emph{AIDS Care}, 19\penalty0 (1):\penalty0 17--19,
  2007{\natexlab{a}}.

\bibitem[Zhang et~al.(2007{\natexlab{b}})Zhang, Lv, Wang, Sun, Zhou, Su, and
  Bi]{zhang2007estimating}
Dapeng Zhang, Fan Lv, Liyan Wang, Liangxian Sun, Jian Zhou, Wenyi Su, and Peng
  Bi.
\newblock Estimating the population of female sex workers in two {C}hinese
  cities on the basis of the {HIV/AIDS} behavioural surveillance approach
  combined with a multiplier method.
\newblock \emph{Sexually Transmitted Infections}, 83\penalty0 (3):\penalty0
  228--231, 2007{\natexlab{b}}.

\bibitem[Abdul-Quader et~al.(2014)Abdul-Quader, Baughman, and
  Hladik]{abdul2014estimating}
Abu~S. Abdul-Quader, Andrew~L. Baughman, and Wolfgang Hladik.
\newblock Estimating the size of key populations: Current status and future
  possibilities.
\newblock \emph{Current Opinion in HIV and AIDS}, 9\penalty0 (2):\penalty0
  107--114, 2014.

\bibitem[Bernard et~al.(1991)Bernard, Johnsen, Killworth, and
  Robinson]{bernard1991estimating}
H~Russell Bernard, Eugene~C Johnsen, Peter~D Killworth, and Scott Robinson.
\newblock Estimating the size of an average personal network and of an event
  subpopulation: Some empirical results.
\newblock \emph{Social Science Research}, 20\penalty0 (2):\penalty0 109--121,
  1991.

\bibitem[Ezoe et~al.(2012)Ezoe, Morooka, Noda, Sabin, and
  Koike]{ezoe2012population}
Satoshi Ezoe, Takeo Morooka, Tatsuya Noda, Miriam~Lewis Sabin, and Soichi
  Koike.
\newblock Population size estimation of men who have sex with men through the
  network scale-up method in {J}apan.
\newblock \emph{PLoS One}, 7\penalty0 (1):\penalty0 e31184, 2012.

\bibitem[Wang et~al.(2015)Wang, Yang, Zhao, Su, Zhao, Chen, Zhang, and
  Zhang]{wang2015application}
Jun Wang, Ying Yang, Wan Zhao, Hualin Su, Yanping Zhao, Yue Chen, Tao Zhang,
  and Tiejun Zhang.
\newblock Application of network scale up method in the estimation of
  population size for men who have sex with men in {S}hanghai, {C}hina.
\newblock \emph{PLoS One}, 10\penalty0 (11):\penalty0 e0143118, 2015.

\bibitem[Sulaberidze et~al.(2016)Sulaberidze, Mirzazadeh, Chikovani, Shengelia,
  Tsereteli, and Gotsadze]{sulaberidze2016population}
Lela Sulaberidze, Ali Mirzazadeh, Ivdity Chikovani, Natia Shengelia, Nino
  Tsereteli, and George Gotsadze.
\newblock Population size estimation of men who have sex with men in tbilisi,
  georgia; multiple methods and triangulation of findings.
\newblock \emph{PLoS One}, 11\penalty0 (2):\penalty0 e0147413, 2016.

\bibitem[Guo et~al.(2013)Guo, Bao, Lin, Wu, Zhang, Hladik, Abdul-Quader,
  Bulterys, Fuller, and Wang]{guo2013estimating}
Wei Guo, Shuilian Bao, Wen Lin, Guohui Wu, Wei Zhang, Wolfgang Hladik, Abu
  Abdul-Quader, Marc Bulterys, Serena Fuller, and Lu~Wang.
\newblock Estimating the size of {HIV} key affected populations in {C}hongqing,
  {C}hina, using the network scale-up method.
\newblock \emph{PLoS One}, 8\penalty0 (8):\penalty0 e71796, 2013.

\bibitem[Maghsoudi et~al.(2014)Maghsoudi, Baneshi, Neydavoodi, and
  Haghdoost]{maghsoudi2014network}
Ahmad Maghsoudi, Mohammad~Reza Baneshi, Mojtaba Neydavoodi, and AliAkbar
  Haghdoost.
\newblock Network scale-up correction factors for population size estimation of
  people who inject drugs and female sex workers in {Iran}.
\newblock \emph{PLoS One}, 9\penalty0 (11):\penalty0 e110917, 2014.

\bibitem[Jami et~al.(2021)Jami, Baneshi, and Nasirian]{jami2021population}
Meysam~Abshenas Jami, Mohammadreza Baneshi, and Maryam Nasirian.
\newblock Population size estimation of high-risk behavior in {Isfahan, Iran}:
  using the network scale-up method in 2018.
\newblock \emph{Journal of Biostatistics and Epidemiology}, 7\penalty0
  (2):\penalty0 120--130, 2021.

\bibitem[Jing et~al.(2018)Jing, Lu, Cui, Yu, and Wang]{jing2018combining}
L~Jing, Q~Lu, Y~Cui, H~Yu, and T~Wang.
\newblock Combining the randomized response technique and the network scale-up
  method to estimate the female sex worker population size: an exploratory
  study.
\newblock \emph{Public Health}, 160:\penalty0 81--86, 2018.

\bibitem[Shelton(2015)]{shelton2015proposed}
Janie~F Shelton.
\newblock Proposed utilization of the network scale-up method to estimate the
  prevalence of trafficked persons.
\newblock In \emph{Forum on Crime and Society}, volume~8, pages 85--94. United
  Nations Publications, 2015.

\bibitem[Killworth et~al.(1998{\natexlab{a}})Killworth, McCarty, Bernard,
  Shelley, and Johnsen]{killworth1998estimation}
Peter~D Killworth, Christopher McCarty, H~Russell Bernard, Gene~Ann Shelley,
  and Eugene~C Johnsen.
\newblock Estimation of seroprevalence, rape, and homelessness in the {U}nited
  {S}tates using a social network approach.
\newblock \emph{Evaluation Review}, 22\penalty0 (2):\penalty0 289--308,
  1998{\natexlab{a}}.

\bibitem[Killworth et~al.(1998{\natexlab{b}})Killworth, Johnsen, McCarty,
  Shelley, and Bernard]{killworth1998social}
Peter~D Killworth, Eugene~C Johnsen, Christopher McCarty, Gene~Ann Shelley, and
  H~Russell Bernard.
\newblock A social network approach to estimating seroprevalence in the {United
  States}.
\newblock \emph{Social Networks}, 20\penalty0 (1):\penalty0 23--50,
  1998{\natexlab{b}}.

\bibitem[UNAIDS and {\relax World Health
  Organization}(2010)]{unaids2010guidelines}
UNAIDS and {\relax World Health Organization}.
\newblock Guidelines on estimating the size of populations most at risk to
  {HIV}.
\newblock Technical report, Geneva, Switzerland, 2010.
\newblock URL
  \url{http://www.unaids.org/en/resources/documents/2011/2011_Estimating_Populations}.

\bibitem[Salganik et~al.(2011)Salganik, Fazito, Bertoni, Abdo, Mello, and
  Bastos]{salganik2011assessing}
Matthew~J. Salganik, Dimitri Fazito, Neilane Bertoni, Alexandre~H. Abdo,
  Maeve~B. Mello, and Francisco~I. Bastos.
\newblock Assessing network scale-up estimates for groups most at risk of
  {HIV/AIDS}: Evidence from a multiple-method study of heavy drug users in
  {Curitiba, Brazil}.
\newblock \emph{American Journal of Epidemiology}, 174\penalty0 (10):\penalty0
  1190, 2011.

\bibitem[Shokoohi et~al.(2012)Shokoohi, Baneshi, and
  Ali-akbar]{shokoohi2012size}
Mostafa Shokoohi, Mohammad~Reza Baneshi, and Haghdoost Ali-akbar.
\newblock Size estimation of groups at high risk of {HIV/AIDS} using network
  scale up in {K}erman, {I}ran.
\newblock \emph{International Journal of Preventive Medicine}, 3\penalty0 (7),
  2012.

\bibitem[Center(2012)]{center2012estimating}
Rwanda~Biomedical Center.
\newblock Estimating the size of key populations at higher risk of {HIV}
  through a household survey ({ESPHS}).
\newblock Technical report, RBC/IHDPC, SPF, UNAIDS and ICF International,
  Calverton, Maryland, USA, 2012.

\bibitem[JafariKhounigh et~al.(2014)JafariKhounigh, Haghdoost, SalariLak,
  Zeinalzadeh, Yousefi-Farkhad, Mohammadzadeh, and
  Holakouie-Naieni]{jafarikhounigh2014size}
Ali JafariKhounigh, Ali~Akbar Haghdoost, Shaker SalariLak, Ali~Hossein
  Zeinalzadeh, Reza Yousefi-Farkhad, Mehdi Mohammadzadeh, and Kourosh
  Holakouie-Naieni.
\newblock Size estimation of most-at-risk groups of hiv/aids using network
  scale-up in {Tabriz}, {Iran}.
\newblock \emph{Journal of Clinical Research \& Governance}, 3\penalty0
  (1):\penalty0 21--26, 2014.

\bibitem[Jing et~al.(2014)Jing, Qu, Yu, Wang, and Cui]{jing2014estimating}
Liwei Jing, Chengyi Qu, Hongmei Yu, Tong Wang, and Yuehua Cui.
\newblock Estimating the sizes of populations at high risk for hiv: a
  comparison study.
\newblock \emph{PLoS One}, 9\penalty0 (4):\penalty0 e95601, 2014.

\bibitem[Teo et~al.(2019)Teo, Prem, Chen, Roellin, Wong, La, and
  Cook]{teo2019estimating}
Alvin Kuo~Jing Teo, Kiesha Prem, Mark~IC Chen, Adrian Roellin, Mee~Lian Wong,
  Hanh~Hao La, and Alex~R Cook.
\newblock Estimating the size of key populations for hiv in singapore using the
  network scale-up method.
\newblock \emph{Sexually Transmitted Infections}, 95\penalty0 (8):\penalty0
  602--607, 2019.

\bibitem[Heydari et~al.(2019)Heydari, Baneshi, Sharifi, Zamanian,
  Haji-Maghsoudi, and Zolala]{heydari2019evaluation}
Zeynab Heydari, Mohammad~Reza Baneshi, Hamid Sharifi, Maryam Zamanian, Saiedeh
  Haji-Maghsoudi, and Farzaneh Zolala.
\newblock Evaluation of the treatment failure ratio in individuals receiving
  methadone maintenance therapy via the network scale up method.
\newblock \emph{International Journal of Drug Policy}, 73:\penalty0 36--41,
  2019.

\bibitem[Haghdoost et~al.(2015)Haghdoost, Baneshi, Haji-Maghsoodi,
  Molavi-Vardanjani, and Mohebbi]{haghdoost2015application}
Ali~Akbar Haghdoost, Mohammad~Reza Baneshi, Saeedeh Haji-Maghsoodi, Hossein
  Molavi-Vardanjani, and Elham Mohebbi.
\newblock Application of a network scale-up method to estimate the size of
  population of breast, ovarian/cervical, prostate and bladder cancers.
\newblock \emph{Asian Pacific Journal of Cancer Prevention}, 16\penalty0
  (8):\penalty0 3273--3277, 2015.

\bibitem[Sajjadi et~al.(2018)Sajjadi, Jorjoran~Shushtari, Shati, Salimi,
  Dejman, Vameghi, Karimi, and Mahmoodi]{sajjadi2018indirect}
Homeira Sajjadi, Zahra Jorjoran~Shushtari, Mohsen Shati, Yahya Salimi, Masoomeh
  Dejman, Meroe Vameghi, Salahedin Karimi, and Zohreh Mahmoodi.
\newblock An indirect estimation of the population size of students with
  high-risk behaviors in select universities of medical sciences: A network
  scale-up study.
\newblock \emph{PLoS One}, 13\penalty0 (5):\penalty0 e0195364, 2018.

\bibitem[Kadushin et~al.(2006)Kadushin, Killworth, Bernard, and
  Beveridge]{kadushin2006scale}
Charles Kadushin, Peter~D. Killworth, H.~Russell Bernard, and Andrew~A.
  Beveridge.
\newblock Scale-up methods as applied to estimates of heroin use.
\newblock \emph{Journal of Drug Issues}, 36\penalty0 (2):\penalty0 417--440,
  2006.

\bibitem[Nikfarjam et~al.(2016)Nikfarjam, Shokoohi, Shahesmaeili, Haghdoost,
  Baneshi, Haji-Maghsoudi, Rastegari, Nasehi, Memaryan, and
  Tarjoman]{nikfarjam2016national}
Ali Nikfarjam, Mostafa Shokoohi, Armita Shahesmaeili, Ali~Akbar Haghdoost,
  Mohammad~Reza Baneshi, Saiedeh Haji-Maghsoudi, Azam Rastegari, Abbas~Ali
  Nasehi, Nadereh Memaryan, and Termeh Tarjoman.
\newblock National population size estimation of illicit drug users through the
  network scale-up method in 2013 in {I}ran.
\newblock \emph{International Journal of Drug Policy}, 31:\penalty0 147--152,
  2016.

\bibitem[Sheikhzadeh et~al.(2016)Sheikhzadeh, Baneshi, Afshari, and
  Haghdoost]{sheikhzadeh2016comparing}
Khodadad Sheikhzadeh, Mohammad~Reza Baneshi, Mahdi Afshari, and Ali~Akbar
  Haghdoost.
\newblock Comparing direct, network scale-up, and proxy respondent methods in
  estimating risky behaviors among collegians.
\newblock \emph{Journal of Substance Use}, 21\penalty0 (1):\penalty0 9--13,
  2016.

\bibitem[Zheng et~al.(2006)Zheng, Salganik, and Gelman]{zheng2006many}
Tian Zheng, Matthew~J. Salganik, and Andrew Gelman.
\newblock How many people do you know in prison? {U}sing overdispersion in
  count data to estimate social structure in networks.
\newblock \emph{Journal of the American Statistical Association}, 101\penalty0
  (474):\penalty0 409--423, 2006.

\bibitem[Bernard et~al.(2001)Bernard, Killworth, Johnsen, Shelley, and
  McCarty]{bernard2001estimating}
H~Russell Bernard, Peter~D Killworth, Eugene~C Johnsen, Gene~A Shelley, and
  Christopher McCarty.
\newblock Estimating the ripple effect of a disaster.
\newblock \emph{Connections}, 24\penalty0 (2):\penalty0 18--22, 2001.

\bibitem[Rastegari et~al.(2014)Rastegari, Baneshi, Haji-Maghsoudi, Nakhaee,
  Eslami, Malekafzali, and Haghdoost]{rastegari2014estimating}
Azam Rastegari, Mohammad~Reza Baneshi, Saiedeh Haji-Maghsoudi, Nowzar Nakhaee,
  Mohammad Eslami, Hossein Malekafzali, and Ali~Akbar Haghdoost.
\newblock Estimating the annual incidence of abortions in {Iran} applying a
  network scale-up approach.
\newblock \emph{Iranian Red Crescent Medical Journal}, 16\penalty0 (10), 2014.

\bibitem[Snidero et~al.(2007)Snidero, Morra, Corradetti, and
  Gregori]{snidero2007use}
Silvia Snidero, Bruno Morra, Roberto Corradetti, and Dario Gregori.
\newblock Use of the scale-up methods in injury prevention research: An
  empirical assessment to the case of choking in children.
\newblock \emph{Social Networks}, 29\penalty0 (4):\penalty0 527--538, 2007.

\bibitem[Yang and Yang(2017)]{yang2017estimating}
Xiaozhao~Yousef Yang and Fenggang Yang.
\newblock Estimating religious populations with the network scale-up method: A
  practical alternative to self-report.
\newblock \emph{Journal for the Scientific Study of Religion}, 56\penalty0
  (4):\penalty0 703--719, 2017.

\bibitem[McCormick et~al.(2010)McCormick, Salganik, and
  Zheng]{mccormick2010many}
Tyler~H McCormick, Matthew~J Salganik, and Tian Zheng.
\newblock How many people do you know?: {E}fficiently estimating personal
  network size.
\newblock \emph{Journal of the American Statistical Association}, 105\penalty0
  (489):\penalty0 59--70, 2010.

\bibitem[Narouee et~al.(2020)Narouee, Shatti, Didevar, and
  Nasehi]{narouee2020estimating}
Sakineh Narouee, Mohsen Shatti, Mahnaz Didevar, and Mahshid Nasehi.
\newblock Estimating social network size using network scale-up method (nsum)
  in {Iranshahr, Sistan and Baluchestan Province, Iran}.
\newblock \emph{Medical journal of the Islamic Republic of Iran}, 34:\penalty0
  35, 2020.

\bibitem[Maltiel et~al.(2015)Maltiel, Raftery, McCormick, and
  Baraff]{maltiel2015estimating}
Rachael Maltiel, Adrian~E Raftery, Tyler~H McCormick, and Aaron~J Baraff.
\newblock Estimating population size using the network scale up method.
\newblock \emph{The Annals of Applied Statistics}, 9\penalty0 (3):\penalty0
  1247, 2015.

\bibitem[Bernard et~al.(2010)Bernard, Hallett, Iovita, Johnsen, Lyerla,
  McCarty, Mahy, Salganik, Saliuk, Scutelniciuc, Shelley, Sirinirund, Weir, and
  Stroup]{bernard2010counting}
H.~R. Bernard, T.~Hallett, A.~Iovita, E.~C. Johnsen, R.~Lyerla, C.~McCarty,
  M.~Mahy, M.~J. Salganik, T.~Saliuk, O.~Scutelniciuc, G.~A. Shelley,
  P.~Sirinirund, S.~Weir, and D.~F. Stroup.
\newblock {Counting hard-to-count populations: The network scale-up method for
  public health}.
\newblock \emph{Sexually Transmitted Infections}, 86\penalty0 (Suppl
  2):\penalty0 ii11--15, 2010.

\bibitem[McCormick(2020)]{mccormick2020network}
Tyler~H McCormick.
\newblock The network scale-up method.
\newblock \emph{The Oxford Handbook of Social Networks}, page 153, 2020.

\bibitem[Laga et~al.(2021)Laga, Bao, and Niu]{laga2021thirty}
Ian Laga, Le~Bao, and Xiaoyue Niu.
\newblock Thirty years of the network scale-up method.
\newblock \emph{Journal of the American Statistical Association}, \penalty0
  (just-accepted):\penalty0 1--33, 2021.

\bibitem[Habecker(2017)]{habecker2017you}
Patrick Habecker.
\newblock \emph{Who Do You Know: Improving and Exploring the Network Scale-Up
  Method}.
\newblock PhD thesis, The University of Nebraska-Lincoln, 2017.

\bibitem[McCormick and Zheng(2007)]{mccormick2007adjusting}
Tyler~H McCormick and Tian Zheng.
\newblock Adjusting for recall bias in “how many x’s do you know?”
  surveys.
\newblock In \emph{Proceedings of the Joint Statistical Meetings}. Citeseer,
  2007.

\bibitem[Habecker et~al.(2015)Habecker, Dombrowski, and
  Khan]{habecker2015improving}
Patrick Habecker, Kirk Dombrowski, and Bilal Khan.
\newblock Improving the network scale-up estimator: Incorporating means of
  sums, recursive back estimation, and sampling weights.
\newblock \emph{PLoS One}, 10\penalty0 (12):\penalty0 e0143406, 2015.

\bibitem[Feehan and Salganik(2016)]{feehan2016generalizing}
Dennis~M. Feehan and Matthew~J. Salganik.
\newblock Estimating the size of hidden populations using the generalized
  network scale-up estimator.
\newblock \emph{Sociological Methodology}, 46\penalty0 (1):\penalty0 153--186,
  2016.

\bibitem[Verdery et~al.(2019)Verdery, Weir, Reynolds, Mulholland, and
  Edwards]{verdery2019estimating}
Ashton~M Verdery, Sharon Weir, Zahra Reynolds, Grace Mulholland, and Jessie~K
  Edwards.
\newblock Estimating hidden population sizes with venue-based sampling:
  Extensions of the generalized network scale-up estimator.
\newblock \emph{Epidemiology (Cambridge, Mass.)}, 30\penalty0 (6):\penalty0
  901, 2019.

\bibitem[McCarty et~al.(2001)McCarty, Killworth, Bernard, Johnsen, and
  Shelley]{mccarty2001comparing}
Christopher McCarty, Peter~D Killworth, H~Russell Bernard, Eugene~C Johnsen,
  and Gene~A Shelley.
\newblock Comparing two methods for estimating network size.
\newblock \emph{Human organization}, 60\penalty0 (1):\penalty0 28--39, 2001.

\bibitem[Erd\H{o}s and R\'{e}nyi(1959)]{erdos1959random}
P~Erd\H{o}s and A~R\'{e}nyi.
\newblock On random graphs {I}.
\newblock \emph{Publicationes Mathematicae}, 6:\penalty0 290--297, 1959.

\bibitem[Diaconis and Janson(2007)]{diaconis2007graph}
Persi Diaconis and Svante Janson.
\newblock Graph limits and exchangeable random graphs.
\newblock \emph{arXiv preprint arXiv:0712.2749}, 2007.

\bibitem[Valliant et~al.(2013)Valliant, Dever, and
  Kreuter]{valliant2013practical}
Richard Valliant, Jill~A Dever, and Frauke Kreuter.
\newblock \emph{Practical tools for designing and weighting survey samples},
  volume~1.
\newblock Springer, 2013.

\bibitem[Robins et~al.(2007)Robins, Pattison, Kalish, and
  Lusher]{robins2007introduction}
Garry Robins, Pip Pattison, Yuval Kalish, and Dean Lusher.
\newblock An introduction to exponential random graph (p*) models for social
  networks.
\newblock \emph{Social networks}, 29\penalty0 (2):\penalty0 173--191, 2007.

\bibitem[Barab{\'a}si and Albert(1999)]{barabasi1999emergence}
Albert-L{\'a}szl{\'o} Barab{\'a}si and R{\'e}ka Albert.
\newblock Emergence of scaling in random networks.
\newblock \emph{Science}, 286\penalty0 (5439):\penalty0 509--512, 1999.

\bibitem[Holland et~al.(1983)Holland, Laskey, and
  Leinhardt]{holland1983stochastic}
Paul~W Holland, Kathryn~Blackmond Laskey, and Samuel Leinhardt.
\newblock Stochastic blockmodels: First steps.
\newblock \emph{Social networks}, 5\penalty0 (2):\penalty0 109--137, 1983.

\bibitem[Watts and Strogatz(1998)]{watts1998collective}
Duncan~J Watts and Steven~H Strogatz.
\newblock Collective dynamics of ‘small-world’ networks.
\newblock \emph{Nature}, 393\penalty0 (6684):\penalty0 440--442, 1998.

\bibitem[Seltman(2012)]{seltman2012approximations}
Howard Seltman.
\newblock Approximations for mean and variance of a ratio.
\newblock \emph{Unpublished note}, 2012.

\end{thebibliography}
\bibliographystyle{unsrtnat}

\newpage
\appendix


\section{Variance approximations}\label{sec:approx}


In order to find the smallest $n$ that satisfies \eqref{eq:error}, we first need to compute the variance of the NSUM estimator in \eqref{eq:nsum}.
Under the marginal degree model in \eqref{eq:marg_bin}, if we assume $d_i$ is fixed, then we have
\begin{equation}\label{eq:NSUM_var}
    \begin{split}
        \Var(\widehat{N}) &= \frac{nM^2}{\left(\sum_{i=1}^n d_i\right)^2} \Var(d_i^u) \\
        &= \frac{nM^2}{\left(\sum_{i=1}^n d_i\right)^2} Np(1-p) \\
        &\approx \frac{nM^2}{(nMp)^2} Np(1-p) \\
        &= \frac{N}{n} \frac{1-p}{p} \enskip .
    \end{split}
\end{equation}
Although early NSUM methods similarly treated these degrees as fixed, \citet{maltiel2015estimating} argues that they should be treated as random.
We show that \eqref{eq:NSUM_var} is a good approximation of the variance even when $d_i$ is random.
In particular, we show it is a conservative estimator, i.e. at least as big as the variance approximations under alternative degree models.
To do so, we begin with a review of two methods for approximating the variance of a ratio of random variables.


\subsection{Taylor expansion for moments}

\label{sec:taylor}

If $Z = f(X, Y) = X/Y$ is a random variable, then we can approximate its moments by taking the expectation of an approximation of $f$ \citep{seltman2012approximations}.
Let $\mathbb{E}[X] = \mu_X$ and $\text{Var}(X) = \sigma_X^2$.
Similarly, let $\mathbb{E}[Y] = \mu_Y$ and $\text{Var}(Y) = \sigma_Y^2$.
Then the first-order approximation of the variance is
\begin{equation}\label{eq:var_1}
    \text{Var}(Z) \approx \frac{\mu_X^2}{\mu_Y^2}\Big(\frac{\sigma_X^2}{\mu_X^2} - 2\frac{\text{Cov}(X, Y)}{\mu_X \mu_Y} + \frac{\sigma_Y^2}{\mu_Y^2}\Big) \enskip .
\end{equation}


Note that this approach exploits the linearity of expectation, but there is no guarantee that inclusion of higher moments will improve the estimates.
For example, the first-order approximation of $\mathbb{E}[Z]$ is
\begin{equation}
    \mathbb{E}[Z] \approx \mu_X / \mu_Y\label{eq:mean_1} \enskip ,
\end{equation}
whereas the second-order approximation is
\begin{equation*}
    \mathbb{E}[Z] \approx \frac{\mu_X}{\mu_Y} - \frac{\text{Cov}(X, Y)}{\mu_Y^2} + \frac{\mu_X \sigma_Y^2}{\mu_Y^3} \enskip ,
\end{equation*}
which, from simulation, is usually not an improvement on \eqref{eq:mean_1}.

\subsection{Variance decomposition}

\label{sec:variance}

For a model that specifies the conditional distribution of $d_i^u ~\vert~ d_i$, or a marginal model whose induced conditonal model is tractable, we can use the law of total variance so that we only need to linearize $Z = 1/Y$:
\begin{equation}\label{eq:EVVE}
    \begin{split}
        \text{Var}(Z) &= \mathbb{E}[\text{Var}(Z ~\vert~ Y)] + \text{Var}(\mathbb{E}[Z ~\vert~ Y]) \\
        &= \mathbb{E}\Big[\frac{\sigma_{X ~\vert~ Y}^2}{Y}\Big] + \text{Var}\Big(\frac{\mu_{X ~\vert~ Y}}{Y}\Big) \\
        &\approx \frac{\mathbb{E}[\sigma_{X ~\vert~ Y}^2]}{\mathbb{E} Y} + \frac{\mathbb{E}[\mu_{X ~\vert~ Y}]^2}{\mu_Y^2}\Big(\frac{\text{Var}(\mu_{X ~\vert~ Y})}{\mathbb{E}[\mu_{X ~\vert~ Y}]^2} + \frac{\sigma_Y^2}{\mu_Y^2}\Big) \\
        &= \frac{\mathbb{E}[\sigma_{X ~\vert~ Y}^2]}{\mu_Y} + \frac{\mu_X^2}{\mu_Y^2}\Big(\frac{\text{Var}(\mu_{X ~\vert~ Y})}{\mu_X^2} + \frac{\sigma_Y^2}{\mu_Y^2}\Big) \\
        &= \frac{\mathbb{E}[\sigma_{X ~\vert~ Y}^2]}{\mu_Y} + \frac{\mu_X^2}{\mu_Y^2}\Big(\frac{\sigma_X^2 - \mathbb{E}[\sigma_{X ~\vert~ Y}^2]}{\mu_X^2} + \frac{\sigma_Y^2}{\mu_Y^2}\Big) \enskip .
    \end{split}
\end{equation}


\subsection{Examples}

One drawback of the model in \eqref{eq:marg_bin} is that it does not guarantee that $d_i^u \leq d_i$.
Instead, by treating $d_i$ as random, it can be shown \citep[Supplementary]{cheng2020estimating} that with general population sampling (from $V \setminus U$), \eqref{eq:marg_bin} implies
\begin{equation*}
    d_i^u ~\vert~ d_i \sim \text{Hypergeometric}(M-1, N, d_i) \enskip .
\end{equation*}

Alternatively, a conditional degree model can be specified to ensure $d_i^u \leq d_i$ via
\begin{equation}\label{eq:cond_bin}
    \begin{split}
        d_i &\sim \text{Bin}(M-1, p) \\
        d_i^u ~\vert~ d_i &\sim \text{Bin}(d_i, p) \enskip .
    \end{split}
\end{equation}

\citet{killworth1998estimation} use the following variant of \eqref{eq:cond_bin} without specifying a model for $d_i$:
\begin{equation}\label{eq:killworth}
    d_i^u ~\vert~ d_i \sim \text{Bin}(d_i, N/M) \enskip .
\end{equation}
Under \eqref{eq:killworth}, \eqref{eq:nsum} is the MLE, which is unbiased with variance $(M-N)/\overline{d}$ \citep{maltiel2015estimating, cheng2020estimating, mccormick2020network}. 

In Table \ref{tab:var_ex}, we provide variance approximations under different degree models.
Recall that our variance estimate in \eqref{eq:NSUM_var} under the model in \eqref{eq:marg_bin} is
\begin{equation*}
    \Var(\widehat{N}) = \frac{N}{n} \frac{1-p}{p} \enskip .
\end{equation*}
This is always greater than the variance estimate for model \eqref{eq:marg_bin} in the first row using linearization and they are equal as $M \to \infty$.
Similarly, as $N \to M$ and $M \to \infty$, the variance estimates using the decomposition for the models in rows two and three converge to twice the estimate in \eqref{eq:NSUM_var}.
For these reasons, the variance in \eqref{eq:NSUM_var} is conservative.

\begin{table}[htb!]
\centering
\resizebox{\textwidth}{!}{%
\begin{tabular}{@{}|l|l|l|@{}}
\toprule
\multicolumn{1}{|c|}{\textbf{Model}}        & \multicolumn{1}{c|}{\textbf{Linearize $X/Y$}}        & \multicolumn{1}{c|}{\textbf{Linearize $1/Y$}} \\ \midrule
\rowcolor[HTML]{EFEFEF} 
$d_i^u \sim \text{Binomial}(N,~p)$               & $\frac{N}{n}\frac{1-p}{p} \Big(1 - \frac{N}{M}\Big)$ & $\approx \frac{N}{n}\frac{1-p}{p}\Big(np(M-N) + 2 + \frac{1}{M} + \frac{N}{M^2}$ \Big)  \\
$d_i^u ~\vert~ d_i \sim \text{Binomial}(d_i,~p)$ & \cellcolor[HTML]{000000}                             & $Mp(1-p)\Big(M + \frac{1}{n}\Big)$ \\
\rowcolor[HTML]{EFEFEF}
$d_i^u ~\vert~ d_i \sim \text{Binomial}(N,~p)$               & \cellcolor[HTML]{000000} & $\frac{N}{n}\frac{1-p}{p} \Big(1 + \frac{N}{M}\Big)$  \\ \bottomrule
\end{tabular}%
}
	\caption{Variance estimates under $d_i \sim \text{Bin}(M-1, p)$ under two approximations: linearizing $X/Y$, explained in Section \ref{sec:taylor}, and linearizing $1/Y$, explained in Section \ref{sec:variance}.}
\label{tab:var_ex}
\end{table}

\section{Derivation of Heuristic \ref{prop:samp_size}}\label{sec:proof}

Recall that we want to find the smallest $n$ such that 
\begin{equation}\label{eq:error_appendix}
    \Pr\left(\vert~\widehat{N}-N\vert~/N < \varepsilon \right) \ge 1-\alpha \enskip .
\end{equation}

By the central limit theorem, when $n$ is large, $\widehat{N}-N$ is approximated by a mean-zero normal distribution whose variance we recall from \eqref{eq:NSUM_var} as
\begin{equation*}
    v_n = \frac{N}{n} \frac{1-p}{p} \enskip .
\end{equation*}

By symmetry, \eqref{eq:error_appendix} becomes
\begin{equation*}
    \Pr(\widehat{N} - N < N\varepsilon) \ge 1 - \alpha/2 \enskip .
\end{equation*}
Dividing by $\sqrt{v_n}$, we have 
\begin{equation*}
    \Pr\left( \frac{\widehat{N}-N}{\sqrt{v_n}} < \frac{N\varepsilon}{\sqrt{v_n}} \right) \ge 1 - \alpha/2 \enskip ,
\end{equation*}
and thus by normality,
\begin{equation*}
    \Phi\left(\frac{N\varepsilon}{\sqrt{v_n}}\right) \ge 1 - \alpha/2 \enskip ,
\end{equation*}
where $\Phi(\cdot)$ is the cumulative distribution function of a standard normal random variable.
It follows that
\begin{equation*}
    \frac{N\varepsilon}{\sqrt{v_n}} \ge \Phi^{-1}(1 - \alpha/2) = z_{\alpha/2} \enskip ,
\end{equation*}
where $z_{\alpha/2}$ is the $1 - \alpha/2$ quantile of the standard normal distribution. Substituting the definition of $v_n$, we find that 
\begin{equation*}
    nN\varepsilon^2 \frac{p}{1-p} \ge z_{\alpha/2}^2 \enskip ,
\end{equation*}
and therefore the minimum sample size is 
\begin{equation*}
    \begin{split}
        n &\ge \frac{z_{\alpha/2}^2}{N\varepsilon^2} \frac{1-p}{p} \\
        &= \frac{z_{\alpha/2}^2}{\varepsilon^2}\frac{1}{Np} \Big(1-p\cdot \frac{M}{M}\Big) \\
        &= \frac{z_{\alpha/2}^2}{\varepsilon^2}\frac{1}{\overline{d_u}} \Big(1 - \frac{\overline{d}}{M}\Big) \\
        &= \frac{z_{\alpha/2}^2}{\varepsilon^2}\frac{1}{q\overline{d}} \Big(1 - \frac{\overline{d}}{M}\Big) \\
        &= \frac{z_{\alpha/2}^2}{\varepsilon^2}\frac{1}{q} \Big(\frac{1}{\overline{d}} - \frac{1}{M}\Big) \enskip .
    \end{split}
\end{equation*}
\begin{equation*}
     \eqno \square
\end{equation*}

\section{Simulations}\label{appendix:simulations}

We assess the relative error and coverage rates using the minimum sample size, by simulation. 
We employ a factorial design that varies the population size ($M = 1000, 5000, 10000$), the prevalence ($q = .01, .03, \ldots,  .51$), the nominal levels ($\alpha = .01, .05, .1, .2$), and the underlying population graph model.
We include i) an Erd\H{o}s-R\'{e}nyi network \citep{erdos1959random}, $G \sim \text{ER}(M,~p = 0.1)$, ii) an exponential random graph \citep{robins2007introduction} with coefficients of $-1$ for edges and triangles, iii) a preferential attachment network \citep{barabasi1999emergence} with power of $1.4$ and $M/20$ new edges at each step, iv) a stochastic block model \citep{holland1983stochastic},
\begin{equation*}
    G \sim \text{SBM}\left(M,~\begin{pmatrix}
        .18 & .15 & .10 \\
        .15 & .12 & .05 \\
        .10 & .05 & .02
    \end{pmatrix} \right) \enskip ,
\end{equation*}
and v) a small-world network \citep{watts1998collective} with a lattice of 50.
We refer to these five random graph models as ER, ERGM, PA, SBM, and small-world, respectively.
Each graph model has the same expected density around $10\%$, which allows us to isolate the effect of the graph topology and assess the robustness to network model mis-specification, i.e. how well our minimum sample size calculation works when the underlying population graph implies degree models that violate \eqref{eq:marg_bin}.

\begin{figure} 
    \centering
    \subfloat{%
        \includegraphics[width=.75\textwidth]{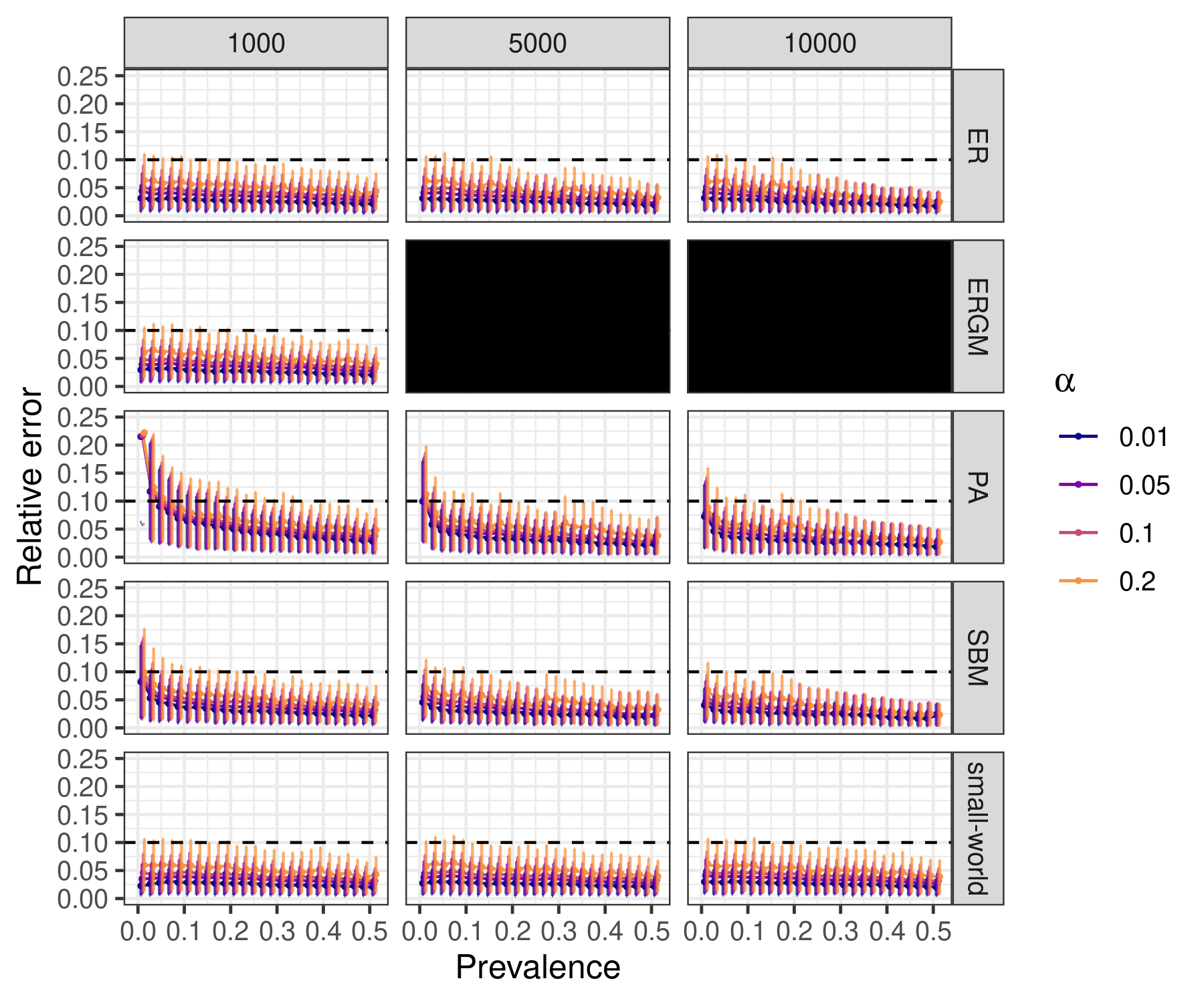}%
    }
    
    \subfloat{%
        \includegraphics[width=.75\textwidth]{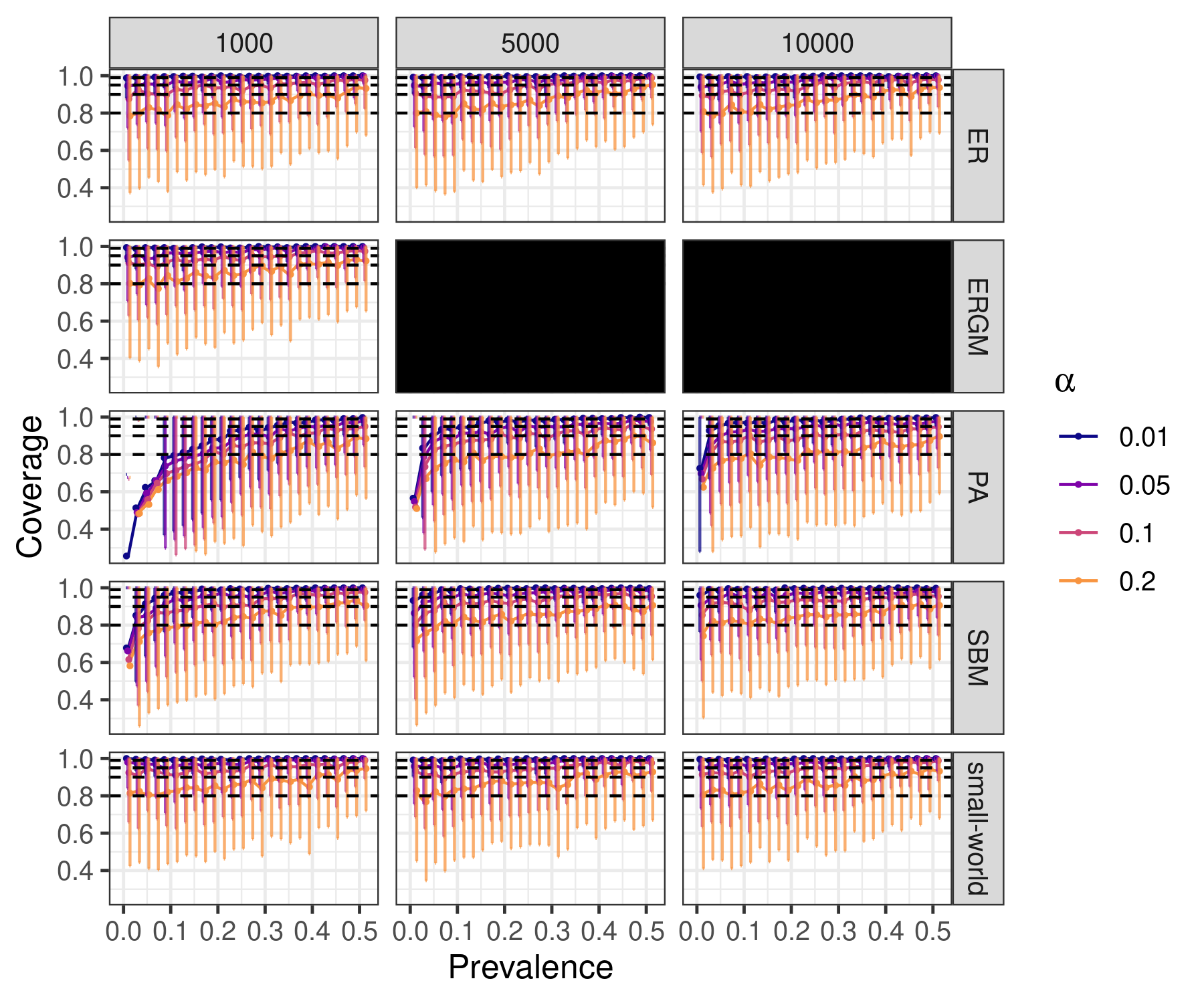}%
    }
    \caption{Relative error of $\widehat{N}$ and confidence interval coverage of the true $N$ over 500 simulated NSUM studies, as a function of prevalence $q=N/M$ and different values of the confidence level $\alpha$. (Top) Average relative error with the dashed line representing the relative error threshold $\varepsilon = 0.1$. (Bottom) Average coverage with the dashed line representing the nominal levels $\alpha \in \{.01, .05, .1, .2\}$. In both plots, error bars are one standard deviation. Black panels indicate infeasibility of computation under ERGM when $M > 1,000$.}
    \label{fig:sim}
\end{figure}

Figure \ref{fig:sim} shows the average relative error and coverage over 500 replicates.
The columns represent the size of the population, $M$, and the rows represent the different (true) underlying population graph model.
In the top plot, the average relative error is below the tolerated level $\varepsilon$ for all values of $M$ and $q$ when the underlying graph model is ER, ERGM, and small-world.
For low prevalence, PA and SBM have average relative errors that exceed the tolerance level, but this is mitigated as prevalence and population size increase.
Results are similar in the bottom plot: the average coverage is conservative for ER, ERGM, and small-world across different nominal levels $\alpha$, whereas it suffers for small populations with low prevalence for PA and SBM.
As expected, the error bars are larger for larger values of $\alpha$.

\begin{figure} 
    \centering
    \subfloat{%
        \includegraphics[width=.75\textwidth]{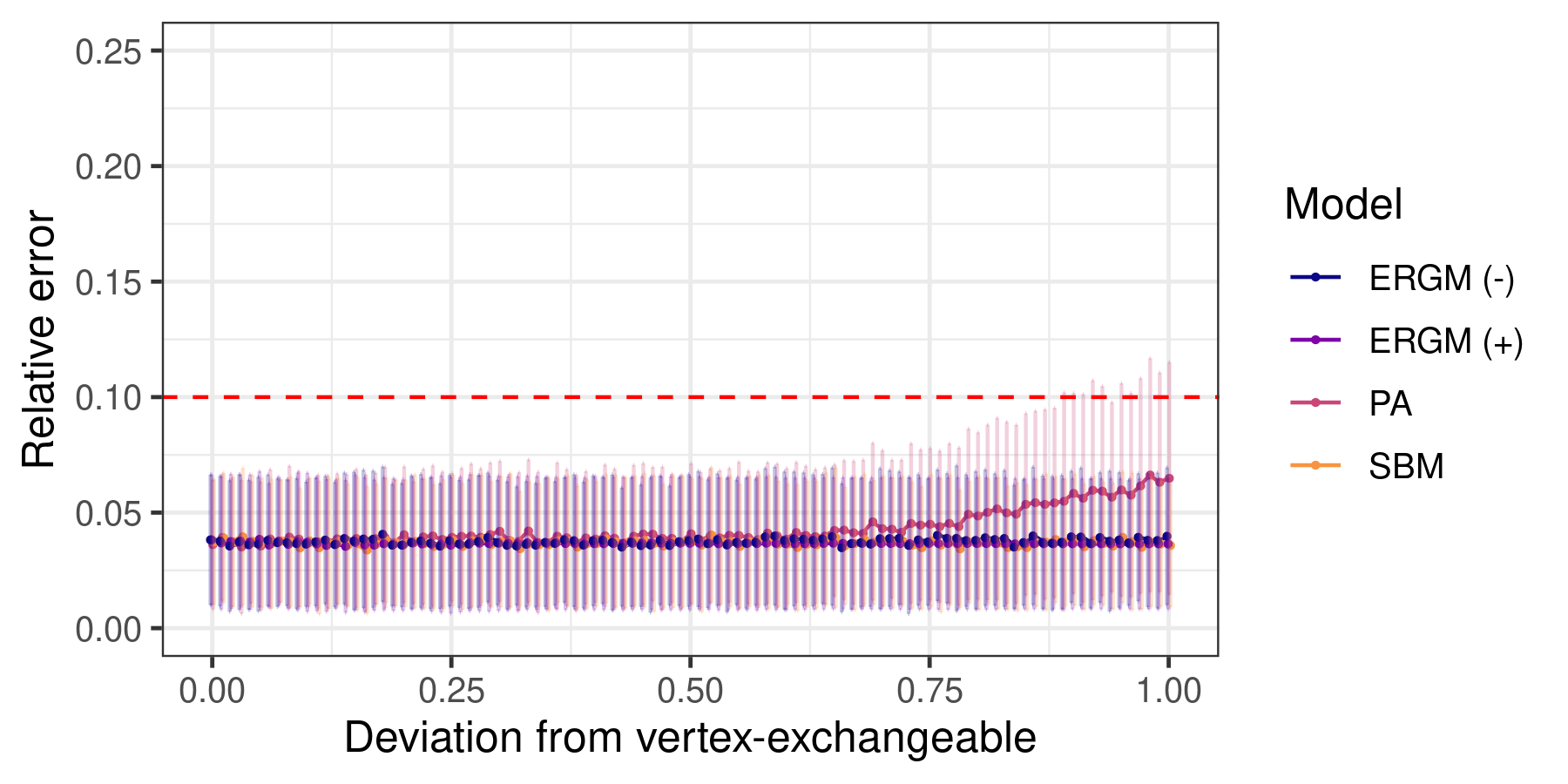}%
    }
    
    \subfloat{%
        \includegraphics[width=.75\textwidth]{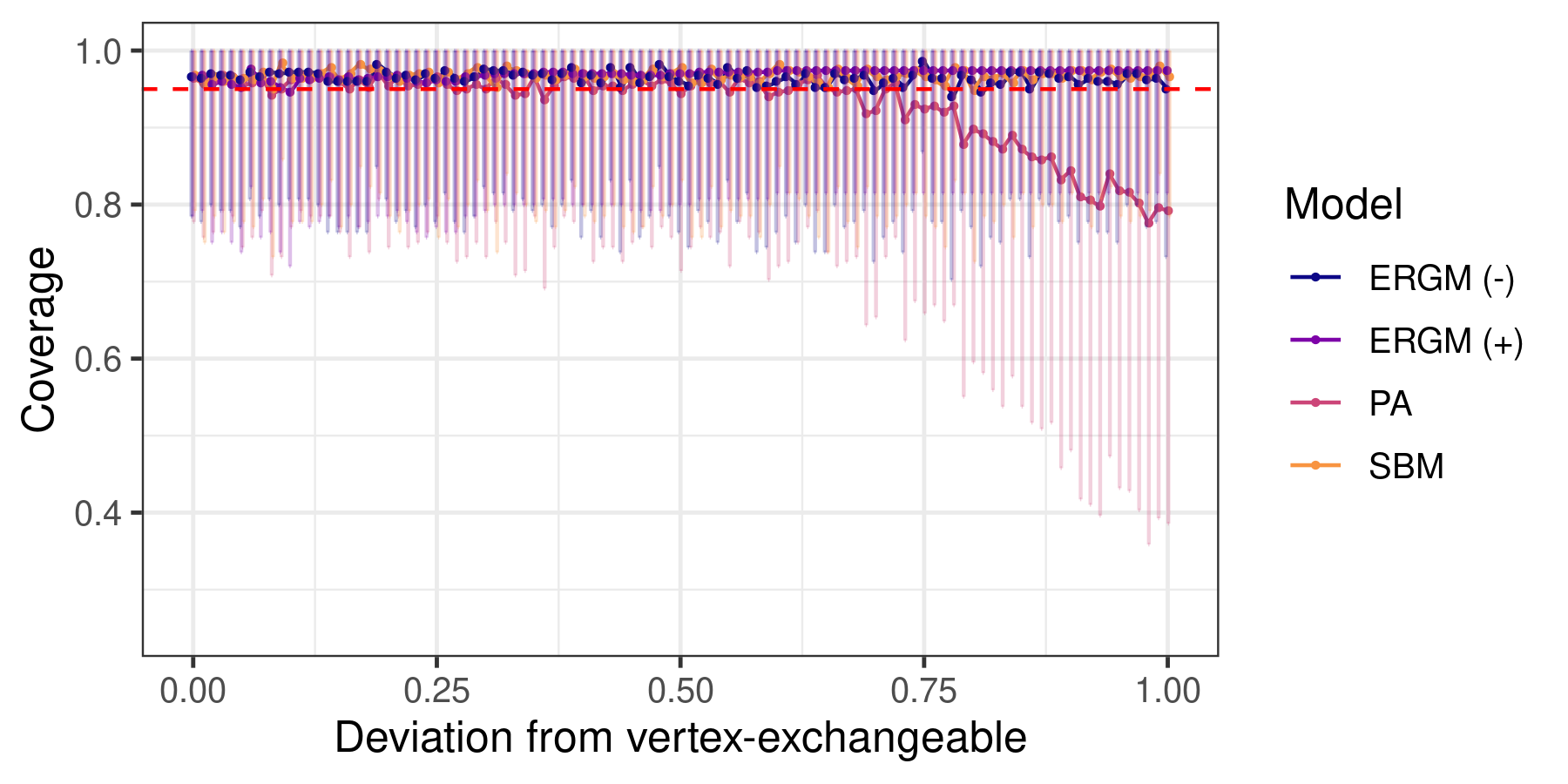}%
    }
    \caption{Relative error of $\widehat{N}$ and confidence interval coverage of the true $N$ over 500 simulated NSUM studies, by random network model.  (Top) Average relative error with the dashed line representing the tolerance $\varepsilon = 0.1$. (Bottom) Average coverage with the dashed line representing the nominal level $\alpha = .05$. In both plots, error bars are one standard deviation.}
    \label{fig:sim.non_exchange}
\end{figure}

Although the minimum sample size works well for all of the graph models, we would like to better understand how the performance changes as the underlying population network gradually deviates away from an Erd\H{o}s-R\'{e}nyi network.
To do so, we consider three models.
First, we have a preferential attachment (PA) network in which $M \cdot (1-\delta)$ of the nodes are an Erd\H{o}s-R\'{e}nyi subgraph.
Second, we have a two-block stochastic block model (SBM) with $p \cdot (1+\delta)$ on the diagonal and $p \cdot (1-\delta)$ off the diagonal.
Finally, we have an exponential random graph model with a triangle coefficient of $\pm \delta$, which we denote ERGM $(\pm)$.
For all of these models, we vary $\delta$ from 0 to 1, and, as before, the average density for all of the networks is fixed at $10\%$.
The results are shown in Figure \ref{fig:sim.non_exchange}.
The error and coverage lie above and below, respectively, the tolerated levels only with extreme deviation from vertex-exchangeability, which further highlights the robustness of our estimator.

\end{document}